\documentclass[letter]{ieice}
\usepackage[dvips]{graphicx}

\field{}
\title{Towards a Decidable LogicWeb via Length-Bounded Derivations}
\authorlist{
\authorentry{Keehang KWON}{m}{labelA}
\authorentry{Daeseong KANG}{m}{labelB}
}
\affiliate[labelA]{The author is with Computer Eng., DongA Univ., Korea.}
\affiliate[labelB]{The author is with Electronics Eng., DongA Univ. The corresponding author.}

\received{2003}{1}{1}
\revised{2003}{1}{1}
\finalreceived{2003}{1}{1}

\newenvironment{numberedlist}
{\begin{list}{\makebox[20pt]{\hss(\arabic{itemno})\enspace}}
             {\usecounter{itemno}\labelwidth 20pt}}{\end{list}}

\newcounter{itemno}

\newcounter{itemno1}

\newcounter{itemno2}

\newcounter{exno}

\newcounter{defno}

\newenvironment{defn}{\refstepcounter{defno}\medskip \noindent {\bf
Definition \thedefno.\ }}{\medskip}

\newcommand{\sep}{\;\vert\;}

\newcommand{\oprove}{\vdash\kern-.6em\lower.7ex\hbox{$\scriptstyle O$}\,}

\newcommand{\Dscr}{{\cal D}}
\newcommand{\Pscr}{{\cal P}}

\newcommand{\Mscr}{{\cal M}}

\newcommand{\pderivation}{{\cal P}\kern -.1em\hbox{\rm -derivation}}
\newcommand{\pderivationl}{{\cal P}\kern -.1em\hbox{\em -derivation}}
\newcommand{\pderivable}{{\cal P}\kern -.1em\hbox{\rm -derivable}}
\newcommand{\pderivablel}{{\cal P}\kern -.1em\hbox{\em -derivable}}
\newcommand{\pderivations}{{\cal P}\kern -.1em\hbox{\rm -derivations}}
\newcommand{\pderivability}{{\cal P}\kern -.1em\hbox{\rm -derivability}}

\newcommand{\all}{\forall}
\newcommand{\some}{\exists}

\newsavebox{\lpartfig}
\newsavebox{\rpartfig}

\newenvironment{exmple}{
 \begingroup \begin{tabbing} \hspace{2em}\= \hspace{3em}\= \hspace{3em}\=
\hspace{3em}\= \hspace{3em}\= \hspace{3em}\= \kill}{
 \end{tabbing}\endgroup}

\newcommand{\lb}{\langle}
\newcommand{\rb}{\rangle}

\newcommand{\Ra}{\supset}  

     {\\* \hspace*{\fill} \end{trivlist}}

\newcommand{\prov}{pv}

\begin{document}
\maketitle
\begin{summary}
LogicWeb has traditionally lacked devices for dealing with intractable queries.
 We address this limitation 
  by adopting length-bounded inference, a form of  approximate reasoning.
A {\it length-bounded} inference is of the form $\prov(\Pscr,G,n)$ which is a success if a query $G$ can be
proved from the web page $\Pscr$ {\it within} $n$ proof steps.
It thus makes LogicWeb decidable and more tractable. During the process, we  propose a novel
module language for logic programming as a device for structuring programs and queries.
\end{summary}

\section{Introduction}\label{sec:intro}

Internet computing is an important modern programming paradigm.
 One successful attempt
towards this direction is LogicWeb\cite{LD96,Kwon08}. LogicWeb is a model of the World Wide Web, where web pages 
are represented as logic programs, and hypertext links represents logical implications
 between these programs.
 Despite much 
attractiveness, LogicWeb (and its relatives such as Semantic Web\cite{Davies}) has 
traditionally lacked elegant devices for dealing with intractable queries.

Dealing with intractable queries is a nontrivial task.
Given the intractability of exact inference in large, distributed webs, it is essential
to consider $approximate$ inference method. 
This paper proposes length-bounded inference which provides approximate inference. 
A {\it length-bounded} inference is of the form $\prov(\Pscr,G,n)$ which is a success if a query $G$ can be
proved from a web page $\Pscr$ {\it within} $n$ proof steps
in a proof system such as the sequent calculus.
Thus the accuracy of inference depends on $n$ and
 the user can choose the degree of accuracy of inference she wants by specifying   a number $n$.


\newcommand{\hweb}{LWeb$^{B}$}

Although it might lose some completeness, there are several advantages with this approach.

\begin{numberedlist}

\item It makes LogicWeb decidable and more tractable, while maintaining good expressibility.
 
\item In the real world, machines do not have  unlimited resources 
and the notion of
length-unbounded derivations is unrealistic.
 The notion of length-bounded derivations is 
a realistic, effective version of its counterpart.

\item Implementing our approach is quite easy and 
requires little overhead to the existing logic programming interpreter.
\end{numberedlist}
\noindent
The notion of length-bounded derivations has been mainly used for analyzing the structures of proofs.
To our knowledge, this is the first time that this notion is used for practical purposes.

 Below we consider LogicWeb based on Horn clause logic  with embedded implications.
This logic extends Horn clauses  by embedded implications
 of the form $D \supset G$ where $D$ is a  Horn clause and $\supset$ is an implication.
This has the following intended semantics: add $D$ to the program in the course of
solving $G$.
The remainder of this paper is structured as follows. We describe \hweb\
 in
the next section. In Section \ref{sec:modules}, we
present some examples of \hweb.
Section~\ref{sec:conc} concludes the paper.

\section{Structuring Programs and Queries via Macros}\label{sec:logic}

Modern languages provide a module language as a device for structuring programs.
We propose a novel device which is  more flexible than traditional module
languages. To be specific,  our language provides a special macro function $/$
which binds a name to a clause or a query. This macro function serves to represent
programs and queries in a concise way. For example, given two macro definitions
$/p\ =\ father(tom,kim)$
and   $/q\ =\ /p$, the notation $/p \land /q$ represents $father(tom,kim) \land father(tom,kim)$.

In contrast to traditional approaches, our module language has the following  characteristics:

\begin{itemize}

\item Our macro languge can be used as a device for structuring queries.
      The difference between macro definition and predicate definition for  structuring queries is 
      that macros definitions are local and temporary. Macro definitions are thus
      not visible to clients.

\item Our macro language is quite flexible, as modules (and queries)  can be freely composed from submodules (and
   subqueries) via diverse logical connectives.

\end{itemize}

Macro definitions are typically processed before the execution  but
in our setting, it is possible to process macros and execute regular programs in an
interleaved fashion.  We adopt this approach below.

In this setting, we describe our language which is an extended  version of Horn clauses
 with  embedded implications. It is described
by $G$-,$D$- and $M$-formulas given by the syntax rules below:
\begin{exmple}
\>$G ::=$ \>   $A \sep /n \sep  G \land  G \sep   D \Ra  G \sep   /n:M \Ra  G  \sep \some x\ G $ \\   \\
\>$D ::=$ \>  $A  \sep  /n \sep G \supset A\ \sep \all x\ D \sep  D \land D$\\
\>$M ::=$ \>  $/n =  G  \sep  /n =  D \sep M \land M$\\
\end{exmple}
\noindent
In the rules above,  $n$ is a name and
$A$  represents an atomic formula.
A $D$-formula  is called a  Horn
 clause with embedded implications. 
 An $M$-formula  is called  macro definitions and $\Mscr$ is a list of $M$-formulas.

Below, $G$-formulas will function as 
queries.  A set of $D$-formulas and a list of $M$-formulas will constitute  a program. 
 We will  present an operational 
semantics for this language. The rules of \hweb\ are formalized by means of what it means to
execute a goal task $G$ from a program $\Pscr$ with respect to $\Mscr$.
 Below the notation $D;\Pscr$ denotes
$\{ D \} \cup \Pscr$ but with the $D$ formula being distinguished
(marked for backchaining). Note that execution  alternates between 
two phases: the goal-reduction phase (one  without a distinguished clause)
and the backchaining phase (one with a distinguished clause).

\begin{defn}\label{def:semantics}
Let $G$ be a goal,  $\Pscr$ be a program and $\Mscr$ be a list macro definitions.
Then the notion of   executing $\lb \Pscr,G\rb$ 
-- $\prov(\Pscr,G)$ -- 
 is defined as follows:
\begin{numberedlist}

\item  $\prov(A;\Pscr,A)$  \% This is a success.

\item    $\prov((G_1\supset A);\Pscr,A$ if 
 $\prov(\Pscr, G_1)$.   \% backchaining

\item    $\prov(\all x D;\Pscr,A)$ if  $\prov([t/x]D;
\Pscr, A)$.    \% instantiation

\item    $\prov(D_1 \land D_2;\Pscr,A)$ if  $\prov(D_1;
\Pscr, A)$.    \%  use $D_1$ to solve $A$.

\item    $\prov(D_1 \land D_2;\Pscr,A)$ if  $\prov(D_2;
\Pscr, A)$.    \% use $D_2$ to solve $A$.

\item    $\prov(/n;\Pscr,A)$ if  $\prov(D;
\Pscr, A)$ and $(/n = D) \in \Mscr$.  \%  we assume it chooses the most recent macro definition.

\item    $\prov(\Pscr,A)$ if  $D \in \Pscr$ and 
$\prov(D;\Pscr, A)$. \%  change to backchaining phase.

\item  $\prov(\Pscr,G_1 \land G_2)$  if 
 $\prov(\Pscr,G_1)$  and
  $\prov(\Pscr,G_2,)$. \% conjunction

\item $\prov(\Pscr,\exists x G_1)$  if $\prov(\Pscr,[t/x]G_1)$.

\item $\prov(\lb \Dscr,\Mscr\rb, D\Ra G_1)$ if  $\prov(\lb \{ D \}\cup \Dscr,\Mscr\rb,G_1)$
\% an implication goal. 

\item $\prov(\lb \Dscr,\Mscr\rb, /n:M \Ra G_1)$ if  $\prov(\lb \{ /n \}\cup \Dscr, M::\Mscr\rb,G_1)$
\% an implication goal. Add new macros to the front of $\Mscr$. Here $::$ is a list constructor.

\item    $\prov(\Pscr, /n)$ if  $(/n = G) \in \Mscr$  and 
$\prov(\Pscr, G)$. \%   a macroed query. We assume it chooses the most recent macro definition.

\end{numberedlist}
\end{defn}

\noindent  

\section{Incorporating Bounded Derivations}\label{sec:logic}

We now incorporate the notion of bounded derivations into our previous language.
This process is quite straightforward.

\begin{defn}\label{def:semantics}
Let $G$ be a goal,  $\Pscr$ be a program, $\Mscr$ be macro definitions  and $m$ be a number.
Then the notion of   executing $\lb \Pscr,G\rb$ within $m$ steps and returning the proof length $n$
-- $\prov(\Pscr,G,m,n)$ -- 
 is defined as follows:
\begin{numberedlist}

\item  $exec(\Pscr,G,m)$ if 
 $\prov(\Pscr,G,m,n)$.   \% call $\prov$

\item  $\prov(A;\Pscr,A,m,1)$ if $ m > 0$. \% This is a success.

\item    $\prov((G_1\supset A);\Pscr,A,m+1,n+1)$ if 
 $\prov(\Pscr, G_1,m,n)$.   \% backchaining

\item    $\prov(\all x D;\Pscr,A,m+1,n+1)$ if  $\prov([t/x]D;
\Pscr, A,m,n)$.    \% instantiation

\item    $\prov(D_1 \land D_2;\Pscr,A,m+1,n+1)$ if  $\prov(D_1;
\Pscr, A,m,n)$.    \%  use $D_1$ to solve $A$.

\item    $\prov(D_1 \land D_2;\Pscr,A,m+1,n+1)$ if  $\prov(D_2;
\Pscr, A,m,n)$.    \% use $D_2$ to solve $A$.

\item    $\prov(/n;\Pscr,A,m+1,n+1)$ if  $\prov(D;
\Pscr, A,m,n)$ and  $(/n = D) \in \Mscr$.  
 \%  a macro definition

\item    $\prov(\Pscr,A,m+1,n+1)$ if  $D \in \Pscr$ and 
$\prov(D;\Pscr, A,m,n)$. \%  change to backchaining phase.

\item  $\prov(\Pscr,G_1 \land G_2,m+1,n_1+n_2+1)$  if 
 $\prov(\Pscr,G_1,m,n_1)$  and
  $\prov(\Pscr,G_2,m-n_1,n_2)$. \% proof length of $G_i$ is $n_i (i=1,2)$.

\item $\prov(\Pscr,\exists x G_1,m+1,n+1)$  if $\prov(\Pscr,[t/x]G_1,m,n)$.

\item $\prov(\lb \Dscr,\Mscr\rb, D\Ra G_1,m+1,n+1)$ if  $\prov(\lb \{ D \}\cup \Dscr,\Mscr\rb, G_1,m,n)$
\% an implication goal. 

\item $\prov(\lb \Dscr,\Mscr\rb, /n:M \Ra G_1)$ if  $\prov(\lb \{ /n \}\cup \Dscr, M::\Mscr\rb,G_1)$
\% an implication goal. Add new macros to the front of $\Mscr$. Here $::$ is a list constructor.

\item    $\prov(\Pscr, /n, m+1,n+1)$ if  $(/n = G) \in \Mscr$  and 
$\prov(\Pscr, G, m,n)$. \%   a macroed query. We assume it chooses the most recent macro definition.

\end{numberedlist}
\end{defn}

\noindent  
In the above rules, choosing  a term $t$  is
a nontrivial task. Fortunately, this poses no problem as
they can be obtained via the well-known unification process. The unification process 
 delays the choices as much as possible until enough information is obtained.
Note that the length of a derivation is based on the standard definition.

\section{\hweb}\label{sec:modules}

In our context, a web page corresponds simply to a set of $M$-formulas
 within a file. 
An example of the use of this construct is provided by the 
following ``lists'' module which contains some basic list-handling rules and
``arcs''  module which contains some arcs between major cities.

\begin{exmple}
 $www.d.com/arcs$. \% file name \\
$/arcs\ =   $ \\
$edge(tokyo,beizing)$. \\
$\vdots$  \\
$edge(paris,london).$
\end{exmple}

\begin{exmple}
 $www.d.com/lists$. \% file name\\
$/lists = /mem \land /app \land /path $.  \\
\%   the path predicate \\
$/path = $\\
$path(X,Y)$ {\rm :-}   $ \ldots $ \\               
\%   the member predicate \\
$/mem = $\\
$memb(X,[X|L]). $\\
$memb(X,[Y|L])$ {\rm :-}  $(neq\ X\ Y)\ \land\  memb(X,L).$\\
\%   the append predicate \\
$/app = $\\
$      append([],L,L). $ \\
$      append([X|L_1],L_2,[X|L_3])$ {\rm :-} $append(L_1,L_2,L_3).$  \\ 
\end{exmple}
\noindent
 These  pages can be made available in specific contexts by explicitly
mentioning the URL via a hyperlink. 

It is well-known that  some queries brings the machine into an infinite sequence of
recursive calls.  Such situations occur frequently in, for example, Prolog \cite{Bratko} with
large programs. 
For example, consider a goal 
{\rm ?-} $www.d.com/lists \Ra www.d.com/arcs \Ra  path(london,boston)$. 
Solving this goal  has the effect
 of adding the rules in $lists, arcs$
to the program before evaluating $path(london,boston)$.
Note that this goal may not be terminating, depending on the implementation of
the $path$ predicate. 

To make this goal more tractable, we may want to use the length-bounded goal such as the one below:\\
{\rm ?- (1000)} $ www.d.com/lists \Ra www.d.com/arcs \Ra  path(london,boston)$. 
This goal obviously terminates, as it must be solved within 1000 proof steps.

\section{Conclusion}\label{sec:conc}

In this paper, we have considered \hweb, a LogicWeb based on an extension to Prolog with  
embedded implications. This extension allows goals of 
the form  $D \Ra G$  where $D$ is a web page and $G$ is a goal.

 \hweb  also adopts the length-bounded inference system in addition to the
 traditional inference system. This feature is 
 particularly useful for  making LogicWeb more tractable.


\section{Acknowledgements}

This work  was supported by Dong-A University Research Fund.

\bibliographystyle{ieicetr}


\end{document}